\documentclass{ws-p8-50x6-00}

\def\Journal#1#2#3#4{{#1} {\bf #2}, #3 (#4)}

\def\NPB{{\em Nucl.~Phys.} B}
\def\NPA{{\em Nucl.~Phys.} A}
\def\PLB{{\em Phys.~Lett.}  B}
\def\PRL{\em Phys.~Rev.~Lett.~}
\def\PRD{{\em Phys.~Rev.} D}

\def\PRep{\em Phys.~Rep.~}

\begin{document}

\title{Signatures of the quark gluon plasma:
\\ a personal overview}

\author{C. Greiner}

\address{Institut f\"ur Theoretische Physik, Universit\"at Giessen,
D-35392 Giessen, Germany
\\E-mail: carsten.greiner@theo.physik.uni-giessen.de}


\maketitle

\abstracts{
In this talk
`{\em Signatures of the Quark-Gluon Plasma}' are being reviewed.
We first discuss, on a no-QGP basis, the two prominent indications
of (a) enhanced strangeness production and of (b)
anomalous $J/\psi $-suppression:
We elaborate in particular on a recent idea of antihyperon
production solely by multi-mesonic reactions.
As a possible source for an enhanced dissociation of
$c \bar{c}$ pairs we summarize the findings
within the 'early'-comover absorption scenario of prehadronic string
excitations.
As an exotic candidate we then finally adress the stochastic formation
of so called disoriented chiral condensates:
An experimentally feasible DCC, if it does exist,
has to be a rare event following an unusual and nontrivial distribution
on an event by event basis.
}

\section{Motivation and Summary}

The prime intention for present ultrarelativistic heavy ion collisions
at CERN and at Brookhaven lies in the possible experimental identification of
the quark gluon plasma (QGP), a theoretically hypothesized
new phase of matter, where quarks and gluons are deliberated
and move freely over an extended, macroscopically large
region. Recently, refering to several different experimental findings
within the Lead Beam Programme at the CERN-SPS,
strong `circumstantial evidence' for the temporal formation of the QGP
has been conjectured\cite{HJ00}.
As a first and principle objection, however, any theoretical predictions
in favor for the QGP can, strictly speaking, only
be regarded as qualitative or as semi-quantitative:
A satisfactory theoretical understanding of either the microscopic
dynamics or of the hadronisation of a hypothetical deconfined phase
is at present not really given.
It is also of scientific importance still to confront the excitement
with further possible criticism.
We therefore want to review, on a {\em no-QGP basis},
where quantitaive predictions are possible, the two most
prominent indications for the QGP:
enhanced strangeness production\cite{KMR86} and anomalous
$J/\psi $-suppression\cite{matsui}, both being proposed already a long time
ago.

The main idea behind the collective enhancement of strangeness
is that the strange (and antistrange) quarks are
thought to be produced more easily and hence also more abundantly
in such a deconfined state as compared to the production via
highly threshold suppressed inelastic hadronic collisions.
The analysis of measured
abundancies of hadronic particles
within thermal models\cite{cleymans} strongly supports the idea of
having established an equilibrated fireball in some late stage of the reaction.
In this respect, especially a nearly fully chemically equilibrated yield
of strange antibaryons, the antihyperons, had originally
been advocated as the appropriate QGP candidate\cite{KMR86}.
Although intriguing, after all this may not be the correct interpretation
of the observed antihyperon yields:
In the following section 2 we will
elaborate in brief on our very recent, yet conservative idea\cite{GL00}
of rapid antihyperon production solely by multi-mesonic reactions like
$n_1\pi + n_2 K \rightarrow \bar{Y}+p $.
This might indeed well explain the observed excess of antihyperons.

A suppression of the
$J/\Psi$ yield in ultra-relativistic heavy-ion collisions
(in comparison to Drell-Yan pairs) is seen as the other
plausible signature because
the strongly bound $J/\Psi$ should dissolve in the QGP
due to color screening.
Indeed, a significant
reduction of the $J/\Psi$ yield when going from proton-nucleus to
nucleus-nucleus collisions has been observed, especially for Pb~+~Pb at
160~GeV/A. Besides the QGP as a possible explanation more conservative
views of possible $J/\Psi $-absorption like on the still incoming nucleons and
also on the produced secondaries, the `comovers', have been envisaged
with rather good success in explaining the data. In these hadronic
scenarios, however, the to be assumed and unknown
annihilation crosssections of the $c\bar{c}$-states on the mesons
are highly debated. As a further and intuitive appealing alternative
we will consider in section 3 within a microscopic simulation
the effect of $c\bar{c}$ dissociation on the individual, highly
excited hadronic strings in
the prehadronic phase of the heavy-ion collision. Such a picture can
in fact be regarded as an 'early'-comover absorption scenario\cite{geiss}.
A satisfactorary agreement with the various data can be achieved
by choosing one phenomenological parameter within a rather plausible range.

\begin{figure}[htb]
\vspace*{1cm}
\begin{center}
\begin{minipage}{9cm}
\epsfxsize=7cm
\epsfbox{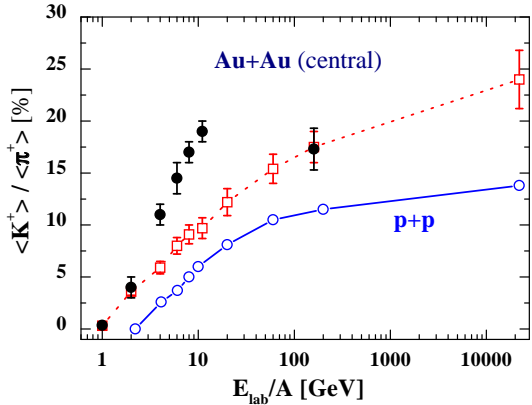}
\end{minipage}
\end{center}
\caption{Calculated $K^+/\pi^+$-ratio\protect\cite{Ge98a} around midrapidity
for central Au+Au reactions (open squares) from SIS to RHIC energies
in comparison to experimental data. For visualization of
the collective strangeness enhancement the corresponding ratio for
elementary p+p collisions (open circles) is also depicted.}
\label{fig:Jochen}
\end{figure}

As a last and  exotic candidate for a direct signature
stemming from the QGP phase transition
(and which can be adressed in RHIC experiments at Brookhaven)
we then summarize our ideas of stochastic formation
of so called disoriented chiral condensates (DCC).
The idea of DCC\cite{DCC} first appeared in a work of Anselm
but it was made widely known due to Bjorken, and Rajagopal and Wilczek.
The spontaneous growth and subsequent
decay of these configurations emerging after a rapid
chiral phase transition from the QGP to the hadronic world
would give rise to large collective fluctuations
in the number of produced low momentum neutral pions compared to charged pions.
In section 4 we briefly summarize our recent findings on
the important question on the
the likeliness of an instability leading potentially to a large DCC yield
of low momentum pions\cite{Xu00}.
Our investigations show that an experimentally feasible DCC, if it does exist
in nature,
has to be a rare event with some finite probability
following a nontrivial and nonpoissonian distribution on an event by event
basis. DCCs could then (only) be revealed experimentally by inspecting
higher order factorial cumulants $\theta _m$ ($m\ge 3$)
in the sampled distribution of low momentum pions.

\section{Strangeness and Antihyperon Production}

\begin{figure}[htp]
\begin{center}
\begin{minipage}{6cm}
\epsfxsize=4cm
\epsfbox{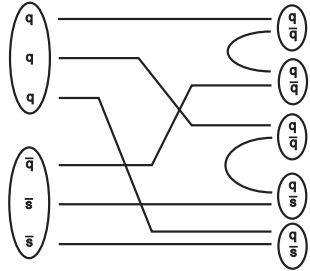}
\end{minipage}
\end{center}
\caption{Schematic picture for 
$\bar{\Xi } + N \leftrightarrow 3 \pi  + 2  K$.}
\label{fig:hyperon}
\end{figure}

Since a relative enhancement of strangeness is observed already
in hadron-hadron collisions  for increasing energy
(see fig.~\ref{fig:Jochen}), which is
certainly not due to any macroscopic or bulk effects, the
to be measured strangeness should be compared
relative to p+p collisions at the same energy.
The arguments for enhanced strangeness
production via the QGP should generally apply already for the most dominant
strange particles, the kaons\cite{KMR86}.
However, nonequilibrium inelastic hadronic reactions
can explain to a very good extent the overall strangeness production
seen experimentally\cite{Ge98}: Within a microscopic transport
simulation an enhancement of the scaled
kaon yield due to hadronic rescattering both with increasing system size
and energy was found.
The outcome for the most dominant strange particles, the
$K^+$-mesons, is summarized in  fig.~\ref{fig:Jochen}.
After the {\em primary} string fragmentation
of intrinsic p-p--collisions the hadronic fireball starts with a
$K^+/\pi^+$ ratio still far below chemical equilibrium with $\approx 6 - 8 \% $
at AGS to SPS energies before the hadronic rescattering starts.
The major amount
of produced strange particles (kaons, antikaons and $\Lambda $s)
at CERN SPS-energies
can then be understood in terms of early and still energetic,
secondary and ternary non-equilibrium interactions.
(At the lower AGS energies, the relative enhancement
factor of $\approx 3$ can not be fully explained within the pure cascade
type calculations \cite{Ge98} without any possible in-medium
modifications of the kaons.)

Still, applying the usual concept of binary collisions
within the transport approaches,
the experimentally observed enhancement of
antihyperons can by far not be explained
by succesive binary (strangeness exchange) reactions\cite{KMR86}.
This fact then gives the strong support for some new exotic mechanism like,
most plausible, the temporary formation of a deconfined and strangeness
saturated new state of matter\cite{HJ00}. As outlined
in the introduction, this might not be the full story\cite{GL00}.
Multimesonic `back-reactions'
(see eg fig.~\ref{fig:hyperon} for a particular illustration)
involving n pions and $n_Y$ kaons of the type
\begin{equation}
\label{antihyp1}
{n}  \pi + n_Y  K \,
\leftrightarrow \,
\bar{Y} + N
\, \, \,
\end{equation}
corresponding to the inverse of the strong binary
baryon-antibaryon annihilation process can easily account for
a fast production of the antihyperon species.
It is the latter annihilation process which dictates the timescale
of how fast the antihyperon densities do approach local
chemical equilibrium with the pions, nucleons and kaons.
A simplified master equation for the number of antihyperons as a function
of time can be written in the most direct form\cite{GL00}
\begin{equation}
\label{masterd}
\frac{d}{dt} \rho _{\bar{Y}} \,  = \,    - \,
\Gamma  _{\bar{Y}}
\left\{
\rho _{\bar{Y}} \, -  \,
\rho ^{eq }_{\bar{Y}}
\right\} \, \, \, ,
\end{equation}
where production due to the multi-mesonic `back-reactions' is hidden in the
second term $\Gamma  _{\bar{Y}} \rho ^{eq }_{\bar{Y}}$.
It is further plausible to assume that the
annihilation crosssections are approximately the same like for $N\bar{p}$
at the same relative momenta, i.e.
$\sigma _{p \bar{Y}\rightarrow n  \pi + n_Y K} \approx 50 $ mb.
The equilibration timescale
$(\Gamma _{\bar{Y}})^{(-1)} \sim
1/ (\sigma _{N \bar{Y}} v _{\bar{Y}N} \rho_B )$
is thus to a good approximation proportional to the
inverse of the baryon density.
Adopting an initial baryon density of approximately 1--2 times
normal nuclear matter density $\rho_0 $ for the initial and thermalized
hadronic fireball, the antihyperons will equilibrate
on a timescale of 1--3 fm/c! This timescale competes
with the expansion timescale of the late hadronic fireball, which
is in the same range or larger. In any case it becomes clear
that these multimesonic, hadronic reactions, contrary to binary
reactions, can explain  most conveniently a sufficiently
fast equilibration before the (so called) chemical freeze-out occurs at
the parameters given by the thermal model analyses\cite{cleymans}.

To be more quantitative some
explicite coupled master equations for an expanding system have to be
considered and are presently pursued.
In addition, one has to invent some clever
strategy to handle such multi-particle `back-reactions'
within the sophisticated transport codes.

\section{$J/\Psi $-suppression via Dissociation by Strings}

\begin{figure}[htb]
\begin{center}
\begin{minipage}{10cm}
\epsfysize=9cm
\epsfbox{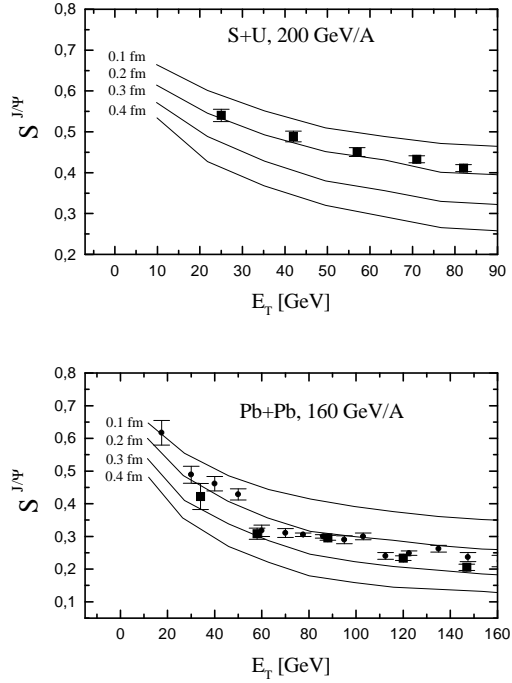}
\end{minipage}
\end{center}
\caption{The $J/\Psi$ survival probability $S^{J/\Psi}$ for S~+~U at 200~
A$\cdot$GeV (upper
part) and Pb~+~Pb at 160~A$\cdot$GeV (lower part) as a function of the
transverse energy $E_T$ in comparison to experimental data
(full squares and circles).
The calculated results are shown for the 
string radii $R_s=$0.1, 0.2, 0.3 and 0.4 fm.
}
\label{jpsi}
\end{figure}

Within a microscopic hadronic transport calculation one can exploit
various assumptions (models) for the $c\bar{c}$ formation and propagation
and also take into account the Drell-Yan process
explicitly.
As one particular scenario we now report on
the effect of $c\bar{c}$ dissociation in
the prehadronic phase of the heavy-ion collision\cite{geiss}.
This is motivated by the fact, that the very early collision phase is 
not described by hadrons but by highly excited strings.
As each individual string carries a lot of internal energy
(to produce the later secondaries) in a small and localized space-time volume
the quarkonia state might get completely dissociated by the
intense color electric field inside a single string \cite{loh}.

In the transport treatment one explicitely follow
the motion of the $c \bar{c}$ pair in the (pre-)hadronic matter
throughout the collision dynamics.
The $c\bar{c}$ pair may now be either destroyed
in collisions with nucleons with a dissociation crosssection of $3-6$ mb
(see the discussion in\cite{geiss}) or by dissociation
on the very energetic prehadronic excitations.
Several hundred of these strings are temporarily
formed during a central Pb-Pb collision
at SPS energies in the early collision stage.
The dynamical evolution of the strings is now
included explicitly~\cite{geiss}.
The fragmentation of the strings into hadrons starts after some
phenomenologically accepted formation time $\tau_f\approx0.8$ fm/c.
For the dissociation we
assume further that a $c\bar{c}$ state immediately
gets broken apart whenever it moves into the region of the color electric
field of a string. In this sense strings are completely 
'black' for $c\bar{c}$ states\cite{loh}.
The field energy density contained in a
string is given by $\sigma / (\pi R_S^2)$,
where $\sigma \approx 1 $GeV/fm  denotes the QCD string tension.
For a string radius $R_S \approx 0.3 \,fm$ one
accordingly has a local high color electric energy density of 
$\approx 4\, GeV/fm^3$, which substantially screens the binding potential
of the charmonium state \cite{matsui}.

The comparison to $J/\Psi$ suppression in nucleus-nucleus collisions
is now performed on an event-by-event basis using the neutral transverse
energy $E_T$ as a trigger as in the experiments.
For p~+~U and a string radius of $R_S=0.4$ fm only 2\% of the $J/\Psi$'s
are absorbed by strings. The absorption is thus dominated, as expected
intuitively, by the $c\bar{c}$-baryon
dissociation on nucleons.
This turns out to be completely different for heavy-ion collisions,
where the absorption on strings becomes a much more important effect.
In Fig.~\ref{jpsi} our results are shown for  S~+~U and Pb~+~Pb as a function of
the transverse energy and for different string radii $R_s$ = 0.1,...,0.4 fm. 
The later serves as the one phenomenological parameter to be
adressed.
A moderate to strong dependence
on the string radius $R_s$ is observed with $R_s \approx$ 0.2-0.3 fm giving
the best fit 
to the experimental data.
For this string radius 40\% of the absorbed
$J/\Psi$'s are dissociated by strings in central collisions 
of Pb~+~Pb.

To summarize, various (hadronic) models can at present achieve
a rather good agreement to all data.
The underlying ideas are to some extent different.
It is, of course, also possible that all of them might attribute
to the suppression of $c\bar{c}$ states, so that the
obliged requirement of a QGP to understand the results of most central
Pb+Pb collisions is not really given.

\section{Stochastic Disoriented Chiral Condensates}

In this last section we will give a brief report on our recent
findings on the stochastic nature of DCC formation and how to
possibly identify their existence experimentally\cite{Xu00}.
This work resulted from an earlier investigation\cite{PRL}, where we
adressed for the first time
the potential likeliness of an instability leading to a
sufficiently large DCC
event during the evolution of a fireball undergoing
a phase transition within the linear $\sigma $-model.

\begin{figure}[htp]
\begin{center}
\begin{minipage}{9cm}
\epsfysize=5cm
\epsfbox{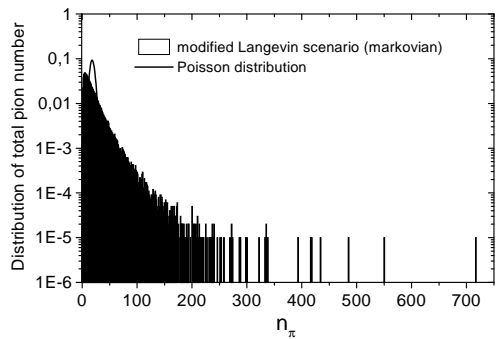}
\end{minipage}
\end{center}
\caption{Statistical distribution $P(n_\pi )$ of the final yield $n_\pi $
in low momentum
pion number of a single DCC for a rapidly expanding
situation (see ref.\protect\cite{Xu00} for details)
compared with a corresponding simple poissonian
distribution.}
\label{DCCfig1}
\end{figure}

The main idea is that the final fluctuations depend critically on the
initial conditions chosen for the evolving chiral order parameter,
thus deciding to some extent whether the system enters temporarily
the instable region during the `roll-down' period
of the order parameter\cite{PRL,Xu00}.
In fact, a semi-classical and dissipative dynamics of the order parameter
and the pionic fields can be obtained by an effective and complex action,
where the interaction with the thermal pions has been integrated out.
(One of the most prominent topics in modern statistical quantum
field theory is to describe the evolution and behavior of the
long wavelength modes at or near thermal equilibrium and also to
understand the non-equilibrium evolution of a phase transition.)
To the end, we have utilized the following stochastic Langevin equations
of motion
for the order parameters $\Phi _a =\frac{1}{V} \int d^3x \,
\phi_a ({\bf x},t)$ in a D-dimensional (`Hubble') expanding volume $V(\tau )$
to describe the evolution of collective pion
and sigma fields\cite{PRL,Xu00}:
\begin{eqnarray}
\ddot{\Phi_0} + \left( \frac{D}{\tau} + \eta \right) \dot{\Phi_0}
+ m_T^2\Phi_0 & = & f_{\pi}m_{\pi}^2 + \xi_0, \nonumber \\
\ddot{\Phi_i} + \left( \frac{D}{\tau} + \eta \right) \dot{\Phi_i}
+ m_T^2\Phi_i & = & \xi_i,
\label{eq1}
\end{eqnarray}
with $\Phi_0=\sigma$ and $\Phi_i=(\pi_1,\pi_2,\pi_3)$
being the chiral meson fields and
$ m_T^2  =  \lambda \left( \Phi_0^2 + \sum_i \Phi_i^2 +
\frac{1}{2} T^2 - f_{\pi}^2 \right) + m_{\pi}^2 $
denotes
the effective transversal (`pionic') masses.
These coupled Langevin equations resemble
in its structure a phenomenological Ginzburg-Landau
description of phase transition.
Aside from a theoretical justification one can  regard the Langevin
equation as a practical tool to study the effect of thermalization
on a subsystem, to sample a large set of possible trajectories
in the evolution, and to address also the question of all thermodynamically
possible initial configurations in a systematic manner.

\begin{figure}[htp]
\begin{center}
\begin{minipage}{9cm}
\epsfysize=5cm
\epsfbox{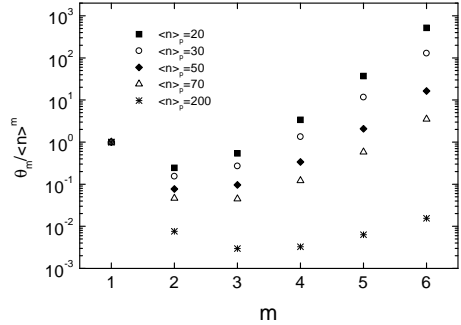}
\end{minipage}
\end{center}
\caption{The reduced factorial cumulants for $m=1$ to 6 for the pion
number distribution of low momentum stemming from
a single emerging DCC (of the previous figure)
and an additional poissonian distributed background pion
source with different mean values $\langle n \rangle_P=20-200$.}
\label{DCCfig2}
\end{figure}

In fig.~\ref{DCCfig1} we show the statistical distribution in the number
of produced long wavelength pions $N_\pi $ out of the evolving chiral
order fields within the DCC domain $V(\tau )$ for one particular set
of parameters\cite{Xu00}. A rather rapid and ($D=$)3-dimensional expansion
has been employed.
(The results majorly depend on how fast the assumed cooling and expansion
proceeds.)
In general one finds that
only for D=3 and sufficiently fast expansion individual unusual strong
fluctuations of the order of 50 - 200 pions might occur,
although the average number $\langle n_\pi \rangle $ of the emerging
long wavelength pions only posesses a moderate and {\em undetectable} value of 5 -20.

In these interesting cases the
final distribution does {\em not} follow a usual Poissonian distribution
(comp. fig.~\ref{DCCfig1}),
which represents a very important outcome of our investigation.
(Critical, dynamical) Fluctuations with a large number of
produced pions are still likely
with some small but finite probability!
Unusual events out of sample contain a multiple in the
number of pions compared to the average.
One should indeed interpret those
particular events as semi-classical `pion bursts' similar to
the mystique Centauro candidates.
This result suggests a very important conclusion:
If DCCs
are being produced, an experimental finding will be a rare event
following a strikingly, nontrivial and nonpoissonian distribution.
A dedicated event-by-event analysis for the experimental
programs (e.g. the STAR TPC at RHIC) is then unalterable.

The further analysis of this unusual 
distribution by means of the cumulant expansion shows that the reduced
higher order factorial cumulants
\\
$\theta_m/<n_{\pi}>^m$ for $m\ge 3$ exhibit
an abnormal, exponentially increasing tendency, as illustrated in
fig.~\ref{DCCfig2}. There an additional incoherent
Poissonian background of (low momentum) pions stemming from other possible
sources has been added.
We advocate that an analysis by
means of the higher order cumulants serves as a new and powerful signature.
In conclusion, the occurence of a rapid chiral phase transition
(and thus DCCs) might then probably only be identified
experimentally by inspecting
higher order facorial cumulants $\theta _m$ ($m\ge 3$)
for taken distributions of low momentum pions.

\section*{Acknowledgments}
The topics reviewed have been done in various collaborations
with T.~Biro, E.~Bratkovskaya, W.~Cassing, J.~Geiss, S.~Leupold,
S.~Loh, U.~Mosel and Z.~Xu.
This work has been supported by BMBF, DFG and GSI Darmstadt.

\end{document}